\documentclass[prb,twocolumn,superscriptaddress,10pt,floatfix]{revtex4-2}
\usepackage{amsmath}
\usepackage{amssymb}
\usepackage{graphicx}
\usepackage{bm}
\usepackage{mathrsfs}
\usepackage{xcolor}
\usepackage{ulem}
\usepackage[colorlinks=true,allcolors=blue]{hyperref}
\usepackage[utf8]{inputenc}

\def\be{\begin{equation}}
\def\ee{\end{equation}}
\def\bea{\begin{eqnarray}}
\def\eea{\end{eqnarray}}
\def\bs{\begin{split}}
\def\es{\end{split}}
\def\ni{\noindent}

\def\mbf{\mathbf}

\def\bi{\begin{itemize}}
\def\ei{\end{itemize}}

\def\a{\alpha}
\def\b{\beta}

\def\p{\partial}
\def\l{\lambda}
\def\e{\epsilon}
\def\vare{\varepsilon}

\def\d{\delta}
\def\o{\omega}
\def\g{\gamma}

\def\f{\frac}
\def\tf{\tfrac}

\begin{document}

\title{Temperature dependence of the charge density from first principles: application to the (222) forbidden reflection in silicon}

\author{Jean Paul Nery}
\altaffiliation{nery.jeanpaul@gmail.com}
\affiliation{Nanomat group, QMAT research unit, and European Theoretical Spectroscopy Facility,  Université de Liège, B5a allée du 6 août, 19,B-4000 Liège, Belgium}
\affiliation{Department of Physics and Astronomy, Stony Brook University, Stony Brook, New York 11794-3800, USA}
\author{Raveena Gupta}
\affiliation{Nanomat group, QMAT research unit, and European Theoretical Spectroscopy Facility,  Université de Liège, B5a allée du 6 août, 19,B-4000 Liège, Belgium}
\author{Olle Hellman}
\affiliation{Weizmann Institute of Science, Rehovot, Israel}
\author{Philip B. Allen}
\affiliation{Department of Physics and Astronomy, Stony Brook University, Stony Brook, New York 11794-3800, USA}
\author{Matthieu J. Verstraete}
\altaffiliation{matthieu.jean.verstraete@gmail.com}
\affiliation{Nanomat group, QMAT research unit, and European Theoretical Spectroscopy Facility,  Université de Liège, B5a allée du 6 août, 19,B-4000 Liège, Belgium}
\affiliation{ITP, Physics Department, Utrecht University, 3508 TA Utrecht, The Netherlands}

\begin{abstract}
Forbidden reflections (FRs) in X-ray diffraction present
inherently weak intensity, and have long been investigated in a variety of
materials, in particular semiconductors like silicon. These reflections serve as
sensitive probes of symmetry breaking,   local strain, impurities, and weak
charge redistribution. Despite extensive experimental work, the theory of the temperature dependence of FRs has typically been addressed through simplified models.
While atomic Debye-Waller factors work well on allowed reflections, their
applicability to the valence charge between atoms -- which determines the intensity of
FRs such as Si (222) -- is questionable, and previous agreement
between theory and experiment relied on ad-hoc Debye-Waller corrections.
In this work we compute the temperature-dependent valence charge density
$\rho(\mathbf{r},T)$ of silicon from first principles, using two methods:
(i) perturbation theory, and (ii) averaging over thermally distorted
supercells in a non-perturbative approach.
The perturbative expression for
the charge density is considerably more demanding than the corresponding one
for electronic energies, since it depends on the wavefunctions themselves and requires an explicit summation over unoccupied bands.
We use an acoustic sum rule
to express the second derivatives of the potential in terms of first
derivatives, making the expression tractable within existing first-principles
frameworks. The (222) FR then follows directly from the Fourier transform of
$\rho(\mathbf{r},T)$, with no ad-hoc factors.
Both methods give similar results, and are in reasonable agreement with
experiment, with thermal expansion having a noticeable effect on the
temperature dependence. The charge density answers a question the measured intensities could not settle: how the valence charge actually redistributes with temperature. Relative to the rigid model, we find more charge in the bonds and less in the core regions, a redistribution that shows up in the intensity as a somewhat weaker temperature dependence of the (222) FR.
\end{abstract}

\maketitle

\section{Introduction}

In perfect crystals, symmetry-imposed selection rules can suppress certain Bragg reflections, rendering them formally ``forbidden'' even though they belong to the reciprocal lattice. These forbidden reflections (FRs) arise when additional symmetries
-- such as glide planes, screw axes, or a multi-atom basis -- cause destructive interference in the structure factor. A classic example is the (222) reflection in silicon: although permitted by the Bravais lattice geometry, it is extinguished due to the two-atom basis in the diamond structure.

FRs have been used for decades as a sensitive probe of subtle symmetry breakings and weak charge redistributions, and X-ray diffraction remains a valuable tool for detecting charge density waves \cite{LeBoloch2016,Blackburn2013}, local strain fields \cite{Shiryaev2020}, impurity effects \cite{Richard2007,Shiryaev2020}, and to study the change of properties or phases with temperature \cite{Isotta2022,Kim2003,Chalise2022}. Resonant X-ray scattering has also been widely employed to study charge and orbital order \cite{Weng2012}, particularly in correlated oxides \cite{Comin2016}. However, while there are many precise and sophisticated experiments \cite{Roberto1974}, theoretical studies remain limited to simplified models \cite{Chelikowsky1974}. This is largely due to the difficulty of computing the temperature dependent charge density, whose perturbative expression is complex and numerically demanding. A common simplification is to use Debye-Waller (DW) factors, which assume a rigid displacement of the charge density. This works well for the usual peaks~\cite{Batterman1962}, since most of the intensity comes from core charge. However, the (222) FR depends solely on the valence charge residing between atoms, where the rigid approximation could be problematic.

The same assumption underlies experimental charge-density analysis, where it is
known as the rigid-pseudoatom or convolution approximation \cite{Coppens1997,Gatti2012}: it
yields a closed form for the structure factor, at the cost of neglecting the
correlation between electronic and nuclear motion. It works well enough in practice that
refinements are routinely carried out only at low temperature, where thermal
smearing is small, and the modeling of thermal motion has consequently
advanced much less than that of the static density itself
\cite{Butkiewicz2025,Hoser2025}. What has not been tested from first principles is whether the approximation holds in the most sensitive regime, for valence charge between atoms and at elevated
temperature.

The (222) FR in silicon was the subject of intense study and careful experimental measurements during the 1970s; these remain the most accurate
measurements of its temperature dependence available.
Ref.~\onlinecite{Colella1966} avoided multiple scattering effects and reported a monotonic decrease of the FR intensity with temperature, in contrast to earlier studies. 	

Neutron scattering, which is sensitive to nuclear displacements but not to the electronic charge, was also used to determine anharmonic contributions to the FR \cite{Roberto1970,Roberto1974}.
Ref.~\onlinecite{Roberto1970} estimated that these account
for less than 5\% of the total signal at room temperature, using a cubic
model \cite{Dawson1967}; the anharmonic term grows with temperature, reaching
about 7\% of the measured (222) amplitude at the highest temperatures
\cite{Roberto1974}. Its effect is opposite to the bond charge contribution, since the atoms spend
less time towards the bond and more towards the ``hole'' on the opposite side.
This also explained the behavior of the even less intense (442) FR \cite{Trucano1972},
whose intensity vanishes around 523~K and increases at higher temperatures.

Ref.~\onlinecite{Phillips1971} pointed out that the same temperature dependence
could equally result from a reduced vibrational amplitude of the bond charge
compensated by a decrease in its magnitude, so that the measured intensity alone
cannot distinguish the two. Using neutron data, Ref.~\onlinecite{Keating1971}
then found the valence-charge scattering to be less temperature dependent than
that of the core, and took this as evidence against the rigid-ion picture.
Combining X-ray and neutron data, Ref.~\onlinecite{Roberto1974} later revised
this, finding instead that the bond contribution in silicon closely follows the
same DW factor as the ion cores, while emphasizing that no definite conclusion
about the bond dynamics could be drawn, and that departures from the DW behavior
do appear in germanium at high temperature.
Ref.~\onlinecite{Chelikowsky1974} then combined a temperature-dependent
scattering factor with an ad-hoc DW factor for the bond derived from an Einstein
model, and reported agreement for both materials. However, the DW factor was
applied at the level of the potential: this yields the response of the full
charge density to the thermally averaged potential but misses the Fan-type
terms, while the DW factor for the bond charge itself remained ad hoc. In
contrast, a consistent treatment would consist in directly evaluating the
Fourier transform of the charge density. The ambiguity arises because the FR is
a single number at each temperature, whereas the underlying object is the full
charge distribution.

At the time, the temperature dependence of electronic properties was not yet understood, and the use of second derivatives (DW) or products of first derivatives (Fan-type terms) was ambiguously applied depending on the problem at hand. This was later clarified through the works of Allen, Heine, and Cardona (AHC)~\cite{Allen1976,Allen1981,Allen1983}, and both terms are normally present in a consistent perturbative approach to second order for the eigenenergies. The same occurs for the charge density, as we will see shortly.
Nowadays, using first-principles calculations, the question of whether the bond charge moves rigidly or whether it decreases relative to the core region can be addressed more directly, without having to rely on approximate models.

In this work we compute the temperature-dependent valence charge density
$\rho(\mathbf{r},T)$ of silicon from first principles, and obtain the (222) FR
directly as its Fourier transform, without ad-hoc factors. This allows us to answer questions that the measured intensities alone
could not settle: whether the bond charge follows the nuclei rigidly, or whether
the amount of charge between the atoms changes with temperature. We focus on silicon, for which the anharmonic contribution to the (222) has
been accurately measured with neutrons \cite{Keating1971,Roberto1974}, so
that the bond contribution can be reliably isolated, and for which standard
functionals provide a robust gapped starting point at all temperatures
considered. First, in
Sec.~\ref{sec:PT}, we derive the perturbative expression for
$\rho(\mathbf{r},T)$, and show how an acoustic sum rule allows it to be
written in terms of first derivatives of the potential alone. Then, in
Sec.~\ref{sec:NP}, we describe a non-perturbative approach based on averaging
the charge density over thermally distorted supercells. In
Sec.~\ref{sec:results} we present and analyze the results, including the
contribution from thermal expansion (TE), and compare the temperature-dependent
charge density obtained with perturbation theory (PT), with the non-perturbative (NP) method, and through the
approximate DW factors. Finally, in Sec.~\ref{sec:conclusions}, we present our
conclusions. An Appendix includes a detailed derivation of the perturbative
expression and other technical aspects. \\

\section{Theoretical background}
\label{sec:theory}

Let $\mbf{k}= 2 \pi \hat{\mbf{n}}/\l$ be the wavevector of an incident X-ray along a direction $\hat{\mbf{n}}$ and wavelength $\l$, and $\mbf{k}'= 2 \pi \hat{\mbf{n}}'/\l$ the wavevector corresponding to a ray scattered along an observation direction $\mbf{\hat{n}}'$. It is well established \cite{AshcroftMermin} that constructive interference and peaks in X-ray diffraction scattering occur when $\mbf{K}:=\mbf{k}-\mbf{k}'$ corresponds to a reciprocal lattice vector, like $(222)$ in silicon. This is usually referred to as coherent scattering.
Taking into account the phase difference between X-rays scattered by electrons in different parts of a crystal \cite{AshcroftMermin}, the scattering amplitude is given by the Fourier transform of the charge density,

\be
S_{\mathbf{K}} = \int d \mathbf{r} \rho(\mathbf{r})e^{i \mathbf{K} \cdot \mathbf{r}},
\label{eq:SK}
\ee

\ni known as the structure factor. The intensity in coherent scattering (see Appendix~\ref{sec:factorization}) can be written as

\be
I = |\langle S_\mbf{K} \rangle |^2
\label{eq:intensity}
\ee

\ni where the brackets indicate the thermal average. 

As we mentioned in the Introduction, if there is more than one atom of the same type in the basis, there can be additional interference that leads to peaks that are much weaker, like the (222) FR of silicon. Assuming that the charge distribution around each atom is the same, then the charge density can be partitioned as $\rho(\mbf{r}) = \sum_i \tilde{\rho}(\mbf{r}-\mbf{R}_i)$, with $\mbf{R}_i$ the position of atom $i$. For example, a common approximation, known as the independent-atom model, is to take $\tilde{\rho}$ as the charge density of the isolated atom \cite{Coppens1997}.
For the diamond lattice this leads to $S_{(h,k,l)}=0$ when $h + k + l = 4p + 2$, as in the case of
(222). This differs from the rigid-pseudoatom model mentioned in the
Introduction, where $\tilde{\rho}$ is not spherical, so that the contributions
do not cancel and the reflection does not vanish.
Although this model is accurate for the core electrons, which have the same spherical symmetry for both atoms in the basis, it is not true for the valence charge, which is tetrahedrally coordinated, and related to the other atom by inversion symmetry around their mid-point. Thus, the full charge density cannot be partitioned in two equal pieces centered in different positions, and there is actually a peak of low intensity, about two orders of magnitude smaller than a standard Bragg peak \cite{Roberto1970}.

Another way to partition the charge density around each atom is with the
ground state (GS) density closest to each atom. Assuming that the charge
around each nucleus moves rigidly with it, the contribution of atom $i$ to
the structure factor acquires a DW factor
$e^{-\tf{1}{2} W_{\mbf{K},i}}$ (see Appendix~\ref{sec:DW} for a derivation), with

\be
W_{\mbf{K},i} := \f{1}{N}\sum_{\mbf{q}s}\f{|\mbf{K} \cdot
\e_{s,i}(\mbf{q})|^2}{2 M_i \o_{\mbf{q}s}} (2 n_{\mbf{q}s}+1),
\ee

\ni where $M_i$ is the mass of atom $i$, $\mbf{q}$ is the phonon wavevector,
$s$ is the phonon mode, $\o_{\mbf{q}s}$ is the frequency of the corresponding
phonon mode, $n_{\mbf{q}s}(T)$ is the Bose-Einstein occupation factor at
frequency $\o_{\mbf{q}s}$ and temperature $T$, $\e_{s,i}(\mbf{q})$ are the
phonon polarization vectors, and $N$ is the number of primitive cells of the
periodic system defined by Born-von K\'arm\'an boundary conditions. Since the two
atoms of the diamond basis are mapped onto each other by the inversion through
the bond center, $W_{\mbf{K},A} = W_{\mbf{K},B} \equiv W_{\mbf{K}}$, and the
Debye-Waller factor factorizes:

\be
S^b_\mbf{K}(T) \propto e^{-\tf{1}{2} W_\mbf{K}(T)}.
\label{eq:DW}
\ee

Applying a DW factor is the standard way of including the temperature
dependence in X-ray diffraction. We improve on this
approximation by using PT and NP approaches, which we now describe in more
detail.\\

\subsection{Perturbation theory}
\label{sec:PT}

The valence charge density 
is given by

\be
\rho(\mathbf{r}) = \f{2}{N \Omega} \sum_{n \mathbf{k}} |\psi_{n \mathbf{k}}(\mathbf{r})|^2
\label{eq:rho}
\ee

\ni where $\psi_{n\mbf{k}}$ is a set of eigenstates, the sum is over the occupied valence orbitals, $n=1,...,4$ (as we mentioned earlier, the contribution from the core states to the FR is 0), and $\Omega$ is the volume of the primitive cell. The wavefunctions are normalized to 1 in the primitive cell.

Using standard PT for the wavefunctions, we obtain an expression that depends on first and second derivatives of the potential. The second derivative term is not routinely implemented in first-principles software packages, and it is common in the AHC approach to express it in terms of first derivatives, making use of the acoustic sum rule (ASR) and rigid ion approximation (RIA)  \cite{Allen1976,Allen1981}. We can do the same here following an analogous procedure for the charge density (see Appendix~\ref{sec:ASR}). In this way, and taking the thermal average, we obtain

\begin{widetext}
\vspace{-0.5cm}
\be
\begin{split}
\langle \delta \rho \rangle  =  & \f{2}{N^2 \Omega}\sum_{\substack{n n' \mathbf{k} \mathbf{q} \\ s\a\b i}}  \psi_{n \mathbf{k}} \psi_{n' \mathbf{k}}^\ast (2 n_{\mathbf{q}s}(T)+1) \times \\
& \times \left\{ \sum_{n''j} V^{i\a,- \mathbf{q}}_{n\mathbf{k},n''\mathbf{k+q}} V^{j\b, \mathbf{q}}_{n''\mathbf{k+q},n'\mathbf{k}}
 \left(\f{1/2}{\Delta \vare^{(2)}_{\substack{n''\mathbf{k+q},n \mathbf{k} \\ n''\mathbf{k+q},n'\mathbf{k}}}} f_{n''} + \f{1}{\Delta \vare^{(2)}_{\substack{ n\mathbf{k},n'\mathbf{k} \hspace{10mm} \\  n\mathbf{k}, n''\mathbf{k+q}}}} f_n - \f{1/2}{\Delta \vare^{(2)}_{n\mathbf{k},n''\mathbf{k+q}}} f_n \d_{nn'} \right) \right. \times \\
& \hspace{3cm} \times \f{\e^\ast_{s,i\a}(\mathbf{q}) \e_{s,j\b}(\mathbf{q})}{2 \o_{\mathbf{q}s} \sqrt{M_i M_j} } \\
& - \left[\sum_{n''j} \f{1}{2} V^{i\a \Gamma}_{n\mathbf{k},n''\mathbf{k}}V^{j\b \Gamma}_{n''\mathbf{k},n'\mathbf{k}} \left( \f{1/2}{\Delta \vare^{(2)}_{ \substack{n''\mathbf{k},n\mathbf{k} \\ n''\mathbf{k},n'\mathbf{k}}}}f_{n''} +  \f{1}{\Delta \vare^{(2)}_{ \substack{n\mathbf{k},n'\mathbf{k} \\ n\mathbf{k},n''\mathbf{k}}}} f_n \right. - \f{1/2}{\Delta \vare^{(2)}_{n\mathbf{k},n'' \mathbf{k}}} f_n \d_{nn'} \right) + (i\a \leftrightarrow j\b) \\
& \left. \left. + \f{1}{2} \f{\overleftarrow{\p}}{\p r_{\b}} \f{V^{i\a \Gamma}_{n\mathbf{k},n'\mathbf{k}}f_{n}}{\vare_{n \mathbf{k}}-\vare_{n'\mathbf{k}}} \rule{0pt}{5.5ex} \right] \times \f{\e^\ast_{s,i\a}(\mathbf{q}) \e_{s,i\b}(\mathbf{q})}{2 \o_{\mathbf{q}s} M_i } \rule{0pt}{5.5ex}  \right\} + \mathrm{c.c.} \\
\end{split}
\label{eq:rho_PT}
\ee
\end{widetext}

\ni where $f_n$ is 1 for valence bands and 0 for unoccupied bands,

\be
\Delta \vare^{(2)}_{\substack{n_1\mathbf{k}_1,n_2 \mathbf{k}_2 \\ n_3\mathbf{k}_3,n_4\mathbf{k}_4}} = (\vare_{n_1\mathbf{k}_1} - \vare_{n_2 \mathbf{k}_2})(\vare_{n_3\mathbf{k}_3}-\vare_{n_4\mathbf{k}_4})
\ee

\ni is a notation to make the expression a little bit more compact, with $\vare_{n\mbf{k}}$ the electronic energy of band $n$ at wavevector $\mbf{k}$ (the band structure). When only one pair of indices is written, both factors are
the same, $\Delta \vare^{(2)}_{n_1\mathbf{k}_1,n_2 \mathbf{k}_2} = (\vare_{n_1\mathbf{k}_1} - \vare_{n_2 \mathbf{k}_2})^2$.
Also,

\be
V^{i\a,\mbf{q}}_{n\mbf{k+q},n'\mbf{k}} = \langle n \mbf{k+q} | \f{\p V}{\p u_{i\a}(\mbf{q})} |n'\mbf{k} \rangle
\ee

\ni and \cite{PonceErratum}

\be
\f{\p V}{\p u_{i\a}(\mbf{q})} = \sum_l \f{\p V}{u_{li\a}} e^{i \mbf{q} \cdot \mbf{R}_l}.
\ee

\ni where $\mbf{R}_l$ is the lattice vector of primitive cell $l$, and $u_{li\a}$ is the displacement of atom $i$ in cell $l$ along Cartesian direction $\a$. The arrow on $\overleftarrow{\p}/\p r_\b$ in Eq.~\eqref{eq:rho_PT} indicates that the derivative acts on the product of wavefunctions to its left.
Using Eq.~\eqref{eq:rho_PT} together with Eqs.~\eqref{eq:SK} and \eqref{eq:intensity}, we obtain the temperature dependence of the forbidden reflection.

This expression is more complex than the well-known AHC expression for the temperature dependence of electronic energies because -- aside from having more terms --
it depends on the wavefunctions, and it involves a sum over electronic wavevectors $\mbf{k}$, in addition to the sum over $\mbf{q}$.
Furthermore, while the
$\delta\psi^{(1)}\delta\psi^{(1)*}$ contribution (see Eqs.~\eqref{eq:deltarho} and \eqref{eq:deltapsi}) can be obtained from
standard Sternheimer solutions, this is not the case for the
$\delta\psi^{(2)}$ term, and an explicit summation over unoccupied bands
remains.

This is challenging, since achieving convergence over electronic properties
requires summing over hundreds of bands \cite{Abreu2022,Ponce2014}. While we can
use symmetries to reduce the number of $\mbf{q}$ for which electron-phonon
matrix elements have to be determined, in general all $\mbf{k}$ must be used
(symmetries can be used for either $\mbf{k}$ or $\mbf{q}$, but not both). The
wavefunctions must therefore be kept for all $\mbf{k}$ and for the $\sim$120
bands needed here, and combining them with the electron-phonon matrix elements
produces heavy intermediate arrays. Great care must be taken in summing the
indices in the correct order, avoiding redundant evaluations in inner loops and
keeping the memory requirements manageable. The numerical parameters are given
in Appendix~\ref{sec:numerical}.

The different objects that enter this expression are calculated using density functional theory (DFT) with the first-principles software package ABINIT \cite{Romero2020}. The wavefunctions and electronic energies are determined from a GS calculation. The phonon frequencies, eigenvectors, and derivatives of the potential are determined using density functional perturbation theory (DFPT) \cite{Gonze1995,Baroni2001}. Additional details can be found in the Appendix.

\subsection{Non-perturbative approach}
\label{sec:NP}

PT obtains derivatives with respect to formally infinitesimal displacements.
However, when displacements are large (high temperatures or light elements), or in strongly anharmonic materials, terms beyond second order in the displacements can become important.

In these cases, NP methods may be more suitable, since they sample the thermal configurations of the system, as opposed to extrapolating properties at large displacements from small displacements. However, NP methods that rely on supercells (SCs) are usually very costly, and may be hard to converge.

Our NP method involves generating stochastic ionic configurations in a SC using harmonic phonons. We consider SCs of size $N_s \times N_s \times N_s$ primitive cells. Let $\e^S_{li\a}$ be the phonon eigenvectors of the SC, with $l$ the index of the primitive cell within the SC, $i$ the atomic index, $\a$ the Cartesian direction, and $S=1,...,6N$ the mode (as opposed to $\mbf{q}s$ for the primitive cell, with $s=1,...,6$), and $n_S$ again the Bose-Einstein factor. It can be seen that the probability distribution is given by \cite{Errea2014}

\be
P \propto \mathrm{exp}\left(- \sum_{ \substack{ lm,ij \\ \a \b, S }}\f{\sqrt{M_{i} M_{j}}\; \o_S }{2 n_S(T) +1} \e^S_{li\a}\e^S_{mj\b} u_{li\a} u_{mj\b}\right).
\ee

We generate the configurations using the TDEP package \cite{TDEP2024}. For each displaced configuration at a given temperature, we obtain the GS charge density using ABINIT, and then average. 
The phonons used to generate the canonical configurations are obtained using DFPT at a fixed lattice constant.
TE is determined with NPT molecular dynamics, which gives better agreement with experiment than the quasi-harmonic approximation (QHA) above 300~K, where the experimental data lie (see Fig.~\ref{fig:QHA} for more details).
We then use TDEP to fit the force constants and generate canonical configurations which account for both the volume and the temperature dependence. This makes the TE case consistent with the sampling method for the constant-lattice case, and also avoids introducing ``distributional anharmonicity'' which was subtracted from the experiments.

\section{Results}
\label{sec:results}

\begin{figure}
\includegraphics[width=0.9\linewidth]{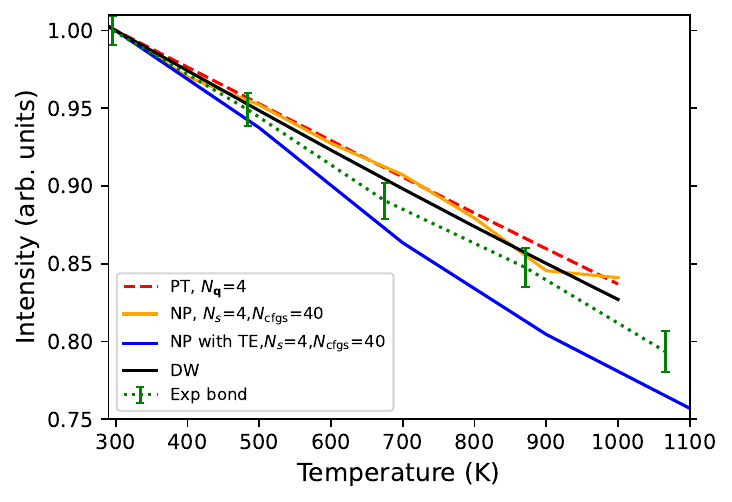}\\
\caption{Temperature dependence of the intensity, normalized at 300~K: PT
(Sec.~\ref{sec:PT}) at a fixed lattice constant (red); the NP method
(Sec.~\ref{sec:NP}) at a fixed lattice constant (orange), NP including TE
(blue); and the rigid DW factor of Eq.~\eqref{eq:DW} (black). Green:
experimental bond intensity from Ref.~\cite{Roberto1974}.}
\label{fig:intensity}
\end{figure}

Using PT Eq.~\eqref{eq:rho_PT} at a fixed lattice constant, we obtain the red curve in Fig.~\ref{fig:intensity}, which is similar to the experimental values in green. The PT curves use $4 \times 4 \times 4$ $\mbf{k}$- and $\mbf{q}$-grids and 120 bands.
Using NP, we obtain the orange curve, which shows a very similar temperature dependence. 
Including thermal expansion in NP increases the temperature dependence, moving the result from slightly above the measured points to slightly below them.
The temperature dependence is weak, so the similarity between PT and NP is not surprising. For comparison, the black curve shows the rigid DW factor of Eq.~\eqref{eq:DW}, which is somewhat steeper than both first-principles methods.

\begin{figure*}[t]
\centering
\setlength{\tabcolsep}{2pt}
\begin{tabular}{cc}
\includegraphics[width=0.47\textwidth]{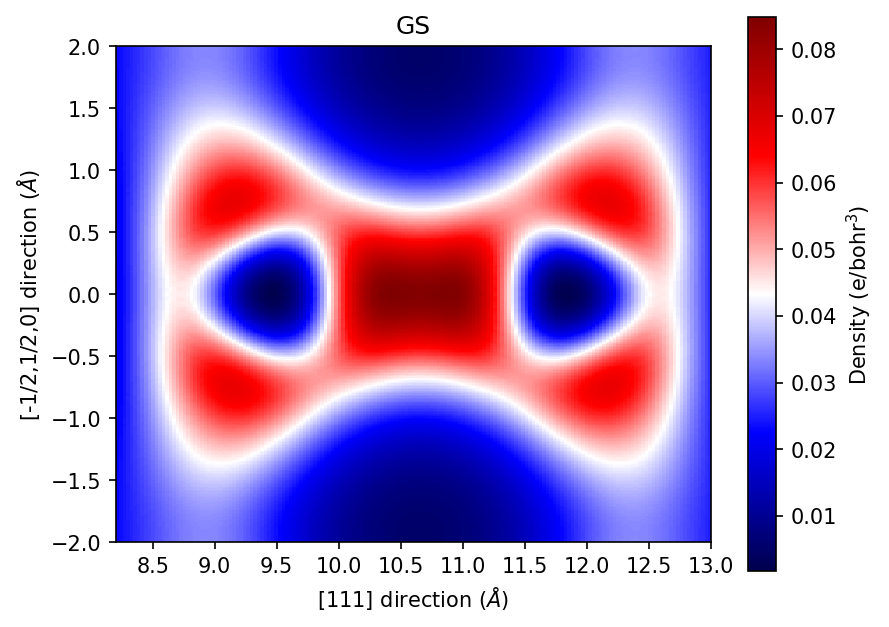} &
\includegraphics[width=0.47\textwidth]{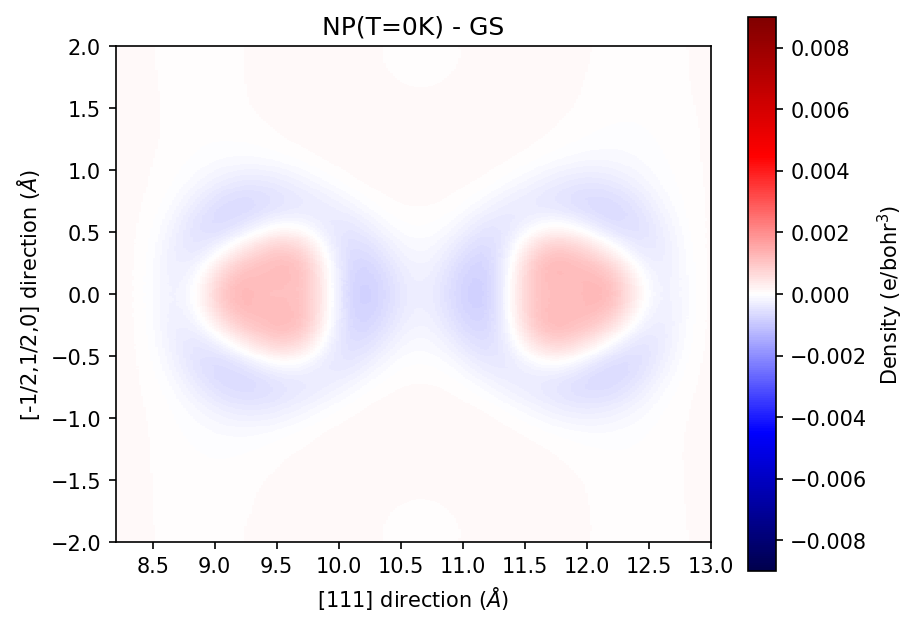} \\
(a) & (b) \\
\includegraphics[width=0.47\textwidth]{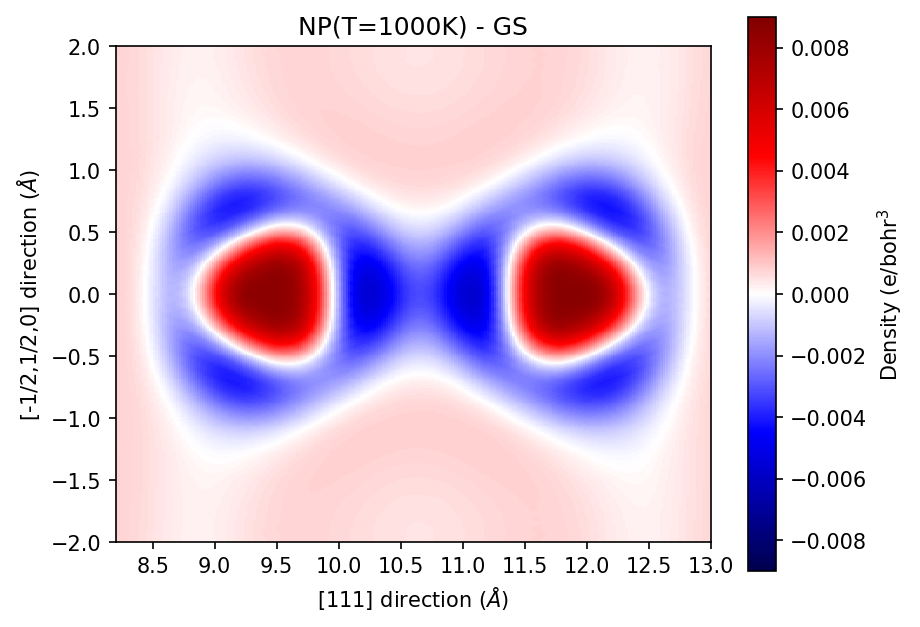} &
\includegraphics[width=0.47\textwidth]{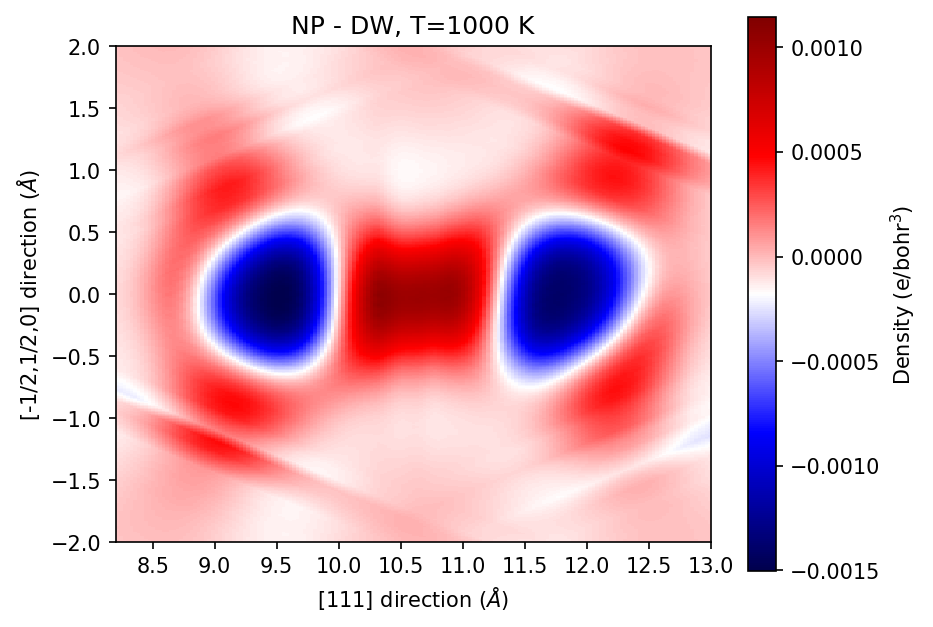} \\
(c) & (d)
\end{tabular}
\caption{Density of (a) the GS and (b) the difference between NP and the GS. (c) NP(1000 K)-GS, (d) NP-DW at 1000 K. The scale is the same in (b) and (c), an order of magnitude smaller than in (a). The magnitude of the density in (b) is an order of magnitude smaller than the scale -- thus the light colors -- which means that the difference between the DFT-GS and adding zero-point effects is very small compared to the difference that arises from higher temperatures (c). To evaluate the DW approximation, in (d) we plot the difference between NP and DW at 1000 K. In NP there is more charge in the bonds, and less in the core region. Additionally, the arbitrary division in DW leads to artifacts.}
\label{fig:density}
\end{figure*}

Convergence of some properties related to electron-phonon coupling, like the band renormalization, requires dense $\mbf{q}$-grids \cite{Ponce2015,Abreu2022}. The FR is less demanding. For NP, $N_s=4$ is already reasonably well converged; for PT, going from the $4\times4\times4$ $\mbf{k}$- and $\mbf{q}$-grids used here to $6\times6\times6$ changes the temperature dependence of Fig.~\ref{fig:intensity}, normalized to 1 at 300~K, by less than 0.5\% over the whole range.
NP includes all orders in the atomic displacements, does not rely on the RIA,
and is much simpler to implement than Eq.~\eqref{eq:rho_PT}, which makes it an
excellent approach to study X-ray diffraction in nonpolar semiconductors and
insulators. The agreement between the two methods indicates in turn that neither
the RIA nor terms beyond second order play a significant role for this
observable in silicon, as was also found for the zero-point renormalization in
diamond~\cite{Ponce2014}.

In order to get further insight into the FR, we analyze the thermalized charge density distribution. A 2D slice of the DFT valence charge density is shown in Fig.~\ref{fig:density}~(a). The two nuclei of the basis, separated by $a/4\,(1,1,1)$ along the horizontal axis, are where the valence charge density is at its lowest (blue). The valence charge is concentrated in between them, at the center of the figure (red); the charge in the other directions corresponds to the bonds with two other neighboring atoms, which lie off the displayed plane. The density at finite temperature looks very similar on this scale, so to visualize the change we plot the difference with respect to the GS at $T=0$~K in (b) and at $T=1000$~K in (c).

To evaluate the DW
approximation, in (d) we plot the difference between NP and DW at 1000~K,
which amounts to about 15\% of the thermal change. Although the overall shape is
similar, the DW construction presents predictable issues: the charge that moves
rigidly must be arbitrarily assigned to one atom or the other, and the
corresponding assignment boundaries produce artifacts in the distribution.
The valence charge therefore does not move rigidly with the nuclei, with more
charge in the bonds and less in the core region than the rigid model predicts. This is not captured by a DW factor at any temperature.

We have addressed the ambiguity raised in the early literature \cite{Phillips1971}. Two pictures were indistinguishable from the measured intensity: a
rigid bond charge with a reduced vibrational amplitude and a bond charge
decreasing in magnitude, but both are incomplete. 
The valence charge redistributes with temperature rather than moving rigidly.
The rigid model somewhat overestimates the temperature dependence of the (222)
relative to both first-principles methods, by an amount of the order of ten
percent of the total change. This is visible not only in the charge density, but in the intensity as well.

\medskip
The structure factor appears to converge quickly in both approaches based on the
calculations performed so far: NP with
the SC size and PT with the $\mbf{k}$- and $\mbf{q}$-grids. This is not usually the case for the temperature
renormalization of $\mbf{k}$-resolved electronic properties \cite{Nery2022},
which refer to individual electronic states. It instead
involves a sum over all electronic wavevectors $\mbf{k}$ in addition to
$\mbf{q}$, so errors from individual states average out and a coarse sampling
can already give a good result.

Although NP is simpler to implement, it requires a separate SC ground state
calculation for each configuration, and even with the apparently fast
convergence, SC calculations are not cheap. PT, now already implemented, works
instead with quantities computed on the primitive cell. So far we have
circumvented the density by evaluating the structure factor directly. Obtaining
the full charge density within PT is more expensive, since the band-resolved
products have to be accumulated on the real-space grid. For future work we are developing a
more memory-efficient implementation, after which the cost should be dominated
by computational time rather than storage.

It remains to be seen what happens in polar materials, where the long-range
electrostatic contribution to the electron-phonon matrix elements diverges as
$1/q$ for the LO modes, and non-adiabatic methods are in principle required
\cite{Ponce2015,Nery2018}. These can be obtained through Green's functions
within PT, while NP is adiabatic by construction and cannot capture them at all, which is an interesting problem for future research.

\section{Conclusions}
\label{sec:conclusions}

We have computed the temperature-dependent valence charge density of silicon
from first principles, using perturbation theory and averaging over thermally
distorted canonical configurations, and obtained the (222) forbidden reflection
directly as its Fourier transform. The two approaches give similar results, and account for the measured temperature dependence without any adjustable parameter, in contrast to a previous treatment which relied on an ad-hoc DW factor for the bond charge.

Our results show that the rigid DW approximation, although it reproduces the
overall trend of the (222) structure factor in silicon, does not reproduce the
detailed redistribution of the valence charge with temperature: the charge
density obtained with the NP method differs from the rigid model, with more charge in the bonds and less in the core region at high temperature.
The charge density resolves this directly rather than through the integrated
intensity, settling a question left open by the original measurements, which
could not distinguish a rigid bond charge of reduced vibrational amplitude from
a bond charge decreasing in magnitude.

More generally, both the perturbative expression derived here and the NP averaging provide a route to
the temperature-dependent charge density that does not rely on rigid-charge
assumptions, and could serve as a predictive tool in other materials where the
valence charge distribution is unknown or hard to access experimentally. It would be particularly interesting to apply it to materials with stronger
anharmonicity, where the rigid assumption is expected to fail more severely,
and to materials with anomalous TE.

\section*{Acknowledgements}
J.P.N. is supported by the European Union under a Marie Sklodowska-Curie Postdoctoral Fellowship, Project GreenNP No. 101151380.
Simulation time was provided by the EuroHPC-JU award EHPC-EXT-2023E02-050 on MareNostrum~5 at the Barcelona Supercomputing Center (BSC), Spain. 
J.P.N. thanks Aloïs Castellano for providing the machine learning potential to obtain the TE.
The authors acknowledge the Fonds de la Recherche Scientifique (FRS-FNRS Belgium) and Fonds Wetenschappelijk Onderzoek (FWO Belgium) for EOS project CONNECT (G.A. 40007563) and 
ARC project DREAMS (G.A. 21/25-11) funded by Federation Wallonie Bruxelles and ULiege.
MJV acknowledges funding by the Dutch Gravitation program
“Materials for the Quantum Age” (QuMat, reg number 024.005.006), financed by the Dutch Ministry of Education, Culture and Science (OCW).

\appendix
\renewcommand{\thefigure}{A\arabic{figure}}
\renewcommand{\theHfigure}{A\arabic{figure}}
\setcounter{figure}{0}

\section{Numerical parameters}
\label{sec:numerical}

\begin{figure}[b]
\includegraphics[width=0.9\linewidth]{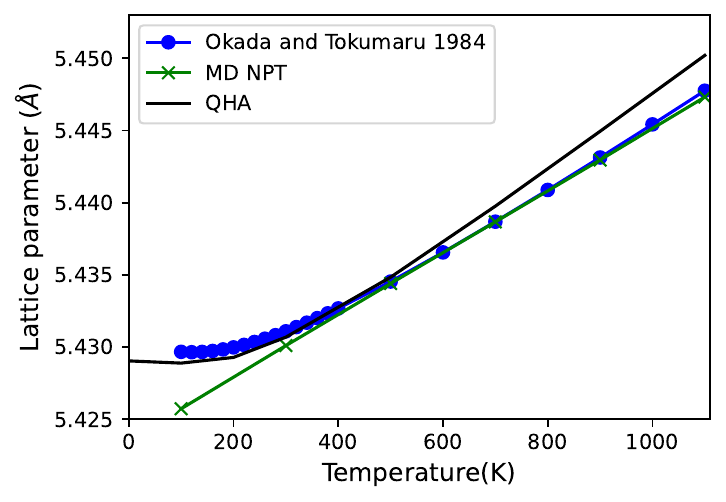}\\
\caption{Lattice parameter as a function of temperature. Here the molecular dynamics (MD) values are shifted by $-0.0075$~\AA ~to visualize how they compare to the other data. The slope of the MD TE agrees better with experiment~\cite{Okada1984} than the one obtained with the QHA.}
\label{fig:QHA}
\end{figure}

All DFT and DFPT calculations use a ONCVPSP norm-conserving pseudopotential from
the PseudoDojo table \cite{PseudoDojo}, with the PBE functional \cite{PBE1996}
and a plane-wave cutoff of 40~Ha. For the NP averages, 40 configurations per temperature are used, and the charge density of each configuration is obtained with the same DFT parameters. The TE is obtained from NPT molecular dynamics with a machine-learning interatomic potential trained on DFT data, which agrees better with experiment than the DFT QHA does, as shown in Fig.~\ref{fig:QHA}.

\section{Derivation}

Starting from Eq.~\eqref{eq:rho}, the variation of the charge density per spin channel to second order in $V$ is

\be
\delta \rho({\mathbf{r}}) = \sum_{ n \mathbf{k}} [\delta \psi^{(2)}_{n \mathbf{k}}(\mathbf{r}) \psi_{n \mathbf{k}}(\mathbf{r})^\ast + \mathrm{c.c.}] + \delta \psi^{(1)}_{n \mathbf{k}} \delta \psi^{\ast (1)}_{n \mathbf{k}}
\label{eq:deltarho}
\ee

\ni A factor $2/\Omega$ is implied, and a $1/N$ is also implied for momentum sums in this Appendix. Let $H = H_0 + V$, with $H_0$ the unperturbed Hamiltonian, and $V$ the perturbing potential due to the displacement of the atoms.

The change of the eigenfunctions to second order is \cite{Sakurai1994}

\begin{widetext}
\be
\begin{split}
\delta \psi_{n \mathbf{k}} = & \sum_{n' \mathbf{k'}} \f{V^{(2)}_{n' \mathbf{k}',n \mathbf{k}}}{\vare_{n \mathbf{k}} - \vare_{n' \mathbf{k}'}} \psi_{n' \mathbf{k}'} +  \sum_{n' \mathbf{k}',n'' \mathbf{k}''} \f{V^{(1)}_{n' \mathbf{k}',n'' \mathbf{k}''} V^{(1)}_{n'' \mathbf{k}'',n \mathbf{k}}}{(\vare_{n \mathbf{k}} - \vare_{n'\mathbf{k}'})(\vare_{n \mathbf{k}}-\vare_{n'' \mathbf{k}''})} \psi_{n' \mathbf{k}'} \\
& - \sum_{n' \mathbf{k}'} \f{V^{(1)}_{n \mathbf{k}, n \mathbf{k}}  V^{(1)}_{n' \mathbf{k}',n \mathbf{k}}}{(\vare_{n \mathbf{k}} - \vare_{n' \mathbf{k}'})^2} \psi_{n' \mathbf{k}'}
-\f{1}{2} \psi_{n \mathbf{k}} \sum_{n' \mathbf{k}'} \f{|V^{(1)}_{n' \mathbf{k}',n \mathbf{k}}|^2}{(\vare_{n \mathbf{k}} - \vare_{n' \mathbf{k}'})^2}
\end{split}
\label{eq:deltapsi}
\ee
\end{widetext}

\ni where the last term, proportional to $\psi_{\mbf{k}n}$, results from the full wavefunction $\psi_{\mbf{k}n} + \d \psi_{\mbf{k}n}$ being normalized to 1 to second order.

The third term of Eq.~\eqref{eq:deltapsi} also vanishes: by momentum conservation, the diagonal matrix element $V_{n\mathbf{k},n\mathbf{k}}$ only involves the $\mathbf{q}=0$ component of the perturbation, and the thermal average restricts the momentum $\mbf{q}'$ of the second matrix element to $\mbf{q}'=-\mbf{q}=0$, so that only $\mathbf{k}'=\mathbf{k}$ survives. With no free sum over $\mathbf{q}$ left to compensate the $1/N$ normalization of the displacements, this term vanishes in the $N \rightarrow \infty$ limit.
 Thus, the adiabatic expression Eq.~\eqref{eq:deltarho} of the charge density to second order in the displacements, $\delta \rho^{(2)}=\sum_{n \mathbf{k}} \delta \psi_{n \mathbf{k}}^{(1)} \delta \psi_{n \mathbf{k}}^{\ast(1)} + [\delta \psi_{n \mathbf{k}}^{(2)} \psi_{n \mathbf{k}}^\ast + \mathrm{c.c.}]$, is:

\begin{widetext}
\be
\begin{split}
\delta \rho^{(2)} & =  \sum_{n \mathbf{k} n' n'' \mathbf{k}'\mathbf{k}''} \f{V^{(1)}_{n \mathbf{k},n'' \mathbf{k}''}  V^{(1)}_{n' \mathbf{k}', n \mathbf{k}}}{(\vare_{n \mathbf{k}} - \vare_{n'' \mathbf{k}''})(\vare_{n \mathbf{k}} - \vare_{n' \mathbf{k}'})} \psi_{n' \mathbf{k}'} \psi_{n'' \mathbf{k}''}^\ast 
 + \left[\sum_{n \mathbf{k} \mathbf{k}' n'} \f{ \tfrac{1}{2} V^{(2)}_{n' \mathbf{k}' ,n \mathbf{k}}}{\vare_{n \mathbf{k}} - \vare_{n' \mathbf{k}'}} \psi_{n' \mathbf{k}'} \psi_{n \mathbf{k}}^\ast \right.\\
& + \sum_{n \mathbf{k} n'n'' \mathbf{k'} \mathbf{k}''} \f{V^{(1)}_{n' \mathbf{k}', n'' \mathbf{k}''} V^{(1)}_{n'' \mathbf{k}'',n \mathbf{k} }}{(\vare_{n \mathbf{k}}-\vare_{n' \mathbf{k}')}(\vare_{n \mathbf{k}} - \vare_{n'' \mathbf{k}''})} \psi_{n' \mathbf{k}'} \psi_{n \textbf{k}}^\ast
 \left. -\f{1}{2} \sum_{n n',\mathbf{k} \mathbf{k}' } \f{|V^{(1)}_{n' \mathbf{k}',n \mathbf{k}}|^2}{(\vare_{n \mathbf{k}} - \vare_{n' \mathbf{k}'})^2}  \psi_{n \mathbf{k}} \psi_{n \textbf{k}}^\ast + \mathrm{c.c.} \right]
\end{split}
\label{delta_rho_1}
\ee
\end{widetext}

\ni where  $V = V^{(1)} + \tfrac{1}{2} V^{(2)}$, with $ V^{(1)}=\tf{\p V}{\p u_{l i \a}} u_{l i \a}$ and $V^{(2)}=\f{\p^2 V}{\p u_{l i \a} \p u_{m j \b}} u_{l i \a} u_{m j \b}$, and the c.c. only affects terms inside the brackets. 

Re-labeling indices (and now actually explicitly writing the complex conjugate of terms 2 and 3 of the previous equation), we can write

\begin{widetext}
\be
\begin{split}
\delta \rho  = & \sum_{\substack{n n' \mathbf{k}\mathbf{k}' \\ li\a, mj\b}}  \psi_{n \mathbf{k}} \psi_{n' \mathbf{k}'}^\ast\times \\
& \times \left\{ \rule{0cm}{1cm} \right. \sum_{n''\mathbf{k}''} V^{li\a}_{n\mathbf{k},n''\mathbf{k}''} V^{mj\b}_{n''\mathbf{k}'',n'\mathbf{k}'} \left[\f{1/2}{\Delta \vare^{(2)}_{\substack{n''\mathbf{k}'',n \mathbf{k} \\ n''\mathbf{k}'',n'\mathbf{k}'}}} f_{n''} + \f{1}{\Delta \vare^{(2)}_{\substack{ n\mathbf{k},n'\mathbf{k}' \\  n\mathbf{k}, n''\mathbf{k}''}}} f_n - \f{1/2}{\Delta \vare^{(2)}_{n\mathbf{k},n''\mathbf{k}''}} f_n \d_{nn'} \d_{\mathbf{k}\mathbf{k}'}\right] \\
& + \left. \f{\tfrac{1}{2}V^{(2)li\a,mj\b}_{n \mathbf{k} ,n' \mathbf{k}'}}{\vare_{n \mathbf{k}} - \vare_{n' \mathbf{k}'}} f_n \rule{0cm}{1cm} \right\} u_{li\a} u_{mj\b} + \mathrm{c.c.}\\
\end{split}
\label{eq:delta_rho}
\ee
\end{widetext}

\ni where we omit the superscript for the first order potential. There is an extra 1/2 in the first term because the c.c. at the end now affects all terms.

Before taking the thermal average, we use an ASR to replace the second derivative, as mentioned in the main text (see next subsection). We obtain

\begin{widetext}
\be
\begin{split}
\delta \rho  = & \sum_{\substack{n n' \mathbf{k}\mathbf{k}' \\ li\a, \b}}  \psi_{n \mathbf{k}} \psi_{n' \mathbf{k}'}^\ast\times \\
& \hspace{-6mm} \times \left\{ \rule{0cm}{1cm} \right. \sum_{\substack{n''\mathbf{k}'' \\ mj}} V^{li\a}_{n\mathbf{k},n''\mathbf{k}''} V^{mj\b}_{n''\mathbf{k}'',n'\mathbf{k}'} \left(\f{1/2}{\Delta \vare^{(2)}_{\substack{n''\mathbf{k}'',n \mathbf{k} \\ n''\mathbf{k}'',n'\mathbf{k}'}}} f_{n''} + \f{1}{\Delta \vare^{(2)}_{\substack{ n\mathbf{k},n'\mathbf{k}' \\  n\mathbf{k}, n''\mathbf{k}''}}} f_n - \f{1/2}{\Delta \vare^{(2)}_{n\mathbf{k},n''\mathbf{k}''}} f_n \d_{nn'} \d_{\mbf{k}\mbf{k}'} \right) u_{li\a} u_{mj\b}  \\
- & \left[ \sum_{\substack{n''\mathbf{k}'' \\ mj}} \f{1}{2} V^{li\a}_{n\mathbf{k},n''\mathbf{k}''} V^{mj\b}_{n''\mathbf{k}'',n'\mathbf{k}'} \left(\f{1/2}{\Delta \vare^{(2)}_{\substack{n''\mathbf{k}'',n \mathbf{k} \\ n''\mathbf{k}'',n'\mathbf{k}'}}} f_{n''} + \f{1}{\Delta \vare^{(2)}_{\substack{ n\mathbf{k},n'\mathbf{k}' \\  n\mathbf{k}, n''\mathbf{k}''}}} f_n - \f{1/2}{\Delta \vare^{(2)}_{n\mathbf{k},n''\mathbf{k}''}} f_n \d_{nn'} \d_{\mbf{k}\mbf{k}'} \right) +(li\a \leftrightarrow mj\b) \right. \\ 
& \left. \left. + \f{1}{2} \f{\overleftarrow{\p}}{\p r_{\b}} \f{V^{li\a}_{n\mathbf{k},n'\mathbf{k}'}f_{n}}{\vare_{n \mathbf{k}}-\vare_{n'\mathbf{k}'}} \right. \Bigg] u_{li\a} u_{li\b} \right\} + \mathrm{c.c.}\\
\end{split}
\label{eq:delta_rho_5}
\ee
\end{widetext}

\ni Writing now the displacements in momentum space as a sum of quantum operators, as done in AHC, taking the thermal average and using momentum conservation \cite{Allen1981}, we get the result of the main text, Eq.~\eqref{eq:rho_PT}.

\section{Acoustic sum rule (ASR)}
\label{sec:ASR}

If under a given distortion $u_{li\a}$ of the SC, both the ionic positions $R_{li\a}$ and observation position $r_\a$ are subject to the same translation $\e_\a$, then the charge density does not change. That is,  $\rho(\mathbf{r},\{ \mathbf{R}\}) = \rho(\mathbf{r} + \pmb{\e},\{ \mathbf{R} + \pmb{\e}\})$. Assuming $\pmb{\e}$ is small, and expanding both arguments in terms of the small displacements $u_{li\a}$ and $\e_\a$, we arrive at

\be
0 = \sum_{li\a} \f{\p^2 \rho}{\p r_\g \p u_{li\a}} u_{li\a} \e_\g + \sum_{li\a} \sum_{mj} \f{\p^2 \rho}{\p u_{li\a} \p u_{mj\g}}u_{li\a} \e_\g 
\ee

\ni Since the displacements $u_{li\a}$ are arbitrary, we can use that only one atom is moving in a given direction, arriving at

\be
0 =  \left(\f{\p^2 \rho}{\p r_\g \p u_{li\a}} + \sum_{mj} \f{\p^2 \rho}{\p u_{li\a} \p u_{mj\g}}\right) u_{li\a} \e_\g. 
\ee

\ni Since $u_{li\a}$ and $\e_\b$ are just numbers, we get that the term inside the parenthesis is 0. This is basically the ASR.

From Eqs.~\eqref{eq:deltarho} and \eqref{eq:deltapsi}, we can see that the first derivative is

\be
\f{\p \rho}{\p u_{li\a}} = \sum_{nn'\mbf{k}\mbf{k}'} \psi_{n\mbf{k}} \psi^\ast_{n'\mbf{k}'} \f{V^{(1)li\a}_{n\mbf{k},n'\mbf{k}'}}{\vare_{n\mbf{k}} - \vare_{n'\mbf{k}'}}f_n + \mathrm{c.c.}
\label{eq:drho_du}
\ee

\ni Now, $\delta \rho = \tf{\p \rho}{\p u_{li\a}} u_{li\a} + \tf{1}{2} \tf{\p^2 \rho}{\p u_{li\a} \p u_{mj\b}} u_{li\a} u_{mj\b}$, and we can write Eq.~\eqref{eq:delta_rho} in this way by symmetrizing in $li\a,mj\b$ the $VV$ term (the $V^{(2)}$ term is already symmetric).
Since the displacements are real, they are unaffected by the complex conjugation and can be factored out of it.

So the ASR is

\begin{widetext}
\be
\begin{split}
0  = &  \sum_{n \mathbf{k} n' \mathbf{k}'} \f{\p}{\p r_{\b}} \psi_{n \mathbf{k}} \psi_{n' \mathbf{k}'}^\ast \f{V^{li\a}_{n\mathbf{k},n'\mathbf{k}'}f_{n}}{\vare_{n \mathbf{k}}-\vare_{n'\mathbf{k}'}} + \mathrm{c.c.}  \\
+ & \sum_{\substack{n n' \mathbf{k}\mathbf{k}' \\ mj}}  \psi_{n \mathbf{k}} \psi_{n' \mathbf{k}'}^\ast\times \\
& \times \left\{ \rule{0cm}{1cm} \right. \sum	 	_{n''\mathbf{k}''} V^{li\a}_{n\mathbf{k},n''\mathbf{k}''} V^{mj\b}_{n''\mathbf{k}'',n'\mathbf{k}'} \left[\f{1/2}{\Delta \vare^{(2)}_{\substack{n''\mathbf{k}'',n \mathbf{k} \\ n''\mathbf{k}'',n'\mathbf{k}'}}} f_{n''} + \f{1}{\Delta \vare^{(2)}_{\substack{ n\mathbf{k},n'\mathbf{k}' \hspace{10mm} \\  n\mathbf{k}, n''\mathbf{k}''}}} f_n - \f{1/2}{\Delta \vare^{(2)}_{n\mathbf{k},n''\mathbf{k}''}} f_n \d_{nn'} \d_{\mathbf{k}\mathbf{k}'}\right] +\\[1em]
& \hspace{1.5cm} + (li\a \leftrightarrow mj\b) \\[1em]
& + \left. \f{V^{(2)li\a,mj\b}_{n \mathbf{k} ,n' \mathbf{k}'}}{\vare_{n \mathbf{k}} - \vare_{n' \mathbf{k}'}} f_n \rule{0cm}{1cm} \right\}  + \mathrm{c.c.}\\
\end{split}
\label{eq:asr}
\ee
\end{widetext}

\ni Using the RIA, we remove the sum over $mj$, write the second derivative in terms of first derivatives, and plug back into Eq.~\eqref{eq:delta_rho}, obtaining Eq.~\eqref{eq:delta_rho_5}. 

\section{Symmetries}

The variation of the density due to phonon vibrations can be written as 

\be
\begin{split}
\delta \rho(\mathbf{r}) & = \sum_{\mathbf{q}\mathbf{k}} \delta \rho(\mathbf{q},\mathbf{k},\mathbf{r})\\
& = \sum_{R}\sum_{\mathbf{q} \in \mathrm{IBZ}} w_\mbf{q} \sum_{\mathbf{k}} \delta \rho(R' \mathbf{q},\mathbf{k},\mathbf{r})\\
\end{split}
\ee

\ni where $w_\mbf{q}=a_\mbf{q}/N_\mathrm{sym}$, $N_\mathrm{sym}$ is the number of symmetries, and $a_\mbf{q}$  is the number of distinct $\mbf{q}$ points in the Brillouin zone related by symmetries to $\mbf{q}$ in the irreducible Brillouin zone (IBZ). For example, for $\mbf{q}=\Gamma$, $a_\mbf{q}=1$, since $R' \mbf{q} = \mbf{q}$ for all symmetries. $R'$ indicates the symmetry operation in reciprocal space, and $R$ in real space. They satisfy $R' = R^{-1,T}$.

Now, we want to remove the rotations acting on $\mbf{q}$, so that electron-phonon matrix elements only have to be calculated for $\mbf{q} \in \mathrm{IBZ}$, by rotating $\mbf{k}$ and $\mbf{r}$ in the opposite direction. 
We can write

\be
\begin{split}
\delta \rho(\mathbf{r}) & = \sum_{R} \sum_{\mathbf{q} \in \mathrm{IBZ}} w_\mbf{q} \sum_{\mathbf{k}} \delta \rho(\mathbf{q},R'^{-1}\mathbf{k},R^{-1}\mathbf{r}) \\
& = \sum_{R} \sum_{\mathbf{q} \in \mathrm{IBZ}} w_\mbf{q} \sum_{\mathbf{k}'} \delta \rho(\mathbf{q},\mathbf{k}',R^{-1}\mathbf{r})
\end{split}
\label{rho_sym}
\ee

\ni Only symmorphic symmetry operations are used, so no fractional translations appear in the positions. 

In the last step we used that rotations are bijective in the BZ, and we are summing over all the BZ, so we are just changing the order of the $\mbf{k}$ points. Since the inverse of a symmetry is also a symmetry, and we are summing over all of them, we can just replace $R^{-1}$ with $R$. 

\section{Order of the sum}

What is a convenient order to do the sums? We can write 
{\small
\be
\begin{split}
\delta \rho(\mbf{r},T) &  \\
 = \sum_{R} \hspace{1mm} & \sum_{\mathbf{q} \in \mathrm{IBZ}} w_\mbf{q} \sum_{\mathbf{k}} \sum_{nn'} \psi_{n\mbf{k}}(\tilde{\mbf{r}}) \psi^\ast_{n'\mbf{k}}(\tilde{\mbf{r}}) P(\mbf{q},\mbf{k},n,n',T)\\
 = \sum_{\mathbf{k}nn'} & \left(\sum_{R} \psi_{n\mbf{k}}(\tilde{\mbf{r}}) \psi^\ast_{n'\mbf{k}}(\tilde{\mbf{r}})\right) \Bigg(\sum_{\mathbf{q} \in \mathrm{IBZ}} w_\mbf{q} P(\mbf{q},\mbf{k},n,n',T)\Bigg)\\
\end{split}
\ee
}

\ni where $\tilde{\mbf{r}} = R \mbf{r}$ includes the rotations, and $P$ includes all of the factors that are not the wavefunctions, including the sum over modes $s$, atoms $i,j$, directions $\a,\b$, and over bands $n''$.
For the derivative term, the first parenthesis with the wavefunctions has a derivative, so an extra index $\b$, and $P$ has an extra Cartesian index $\b$ as well (corresponding to one of the indices of the eigenvectors). We proceed in two ways to obtain the structure factor.

\subsection{Real space}

In this case, we actually obtain the wavefunction in real space from ABINIT's output, which contains the plane-wave coefficients $c_{n,\mbf{k}+\mbf{G}}$, 

\be
\psi_{n\mbf{k}} = \sum_\mbf{G} c_{n,\mbf{k}+\mbf{G}} e^{i (\mbf{k}+\mbf{G}) \cdot \mbf{r}}.
\label{eq:psi_G}
\ee

\ni We multiply the wavefunctions and do the sum over rotations to obtain the first parenthesis. To construct the $P$ term efficiently and avoid building unnecessarily large memory arrays, the calculation evaluates the sum over $\mathbf{q}$ iteratively. For a given $\mathbf{q}$-point, the phonon polarization vectors are first normalized by the square root of the atomic masses and phonon frequencies. These normalized eigenvectors are then contracted with the electron-phonon matrix elements and combined with the corresponding energy denominators to assemble the intermediate components. These components are accumulated into a running total as the integration over $\mathbf{q}$ proceeds. Finally, the temperature dependence is introduced at the very end by multiplying the accumulated values by the Bose-Einstein thermal occupation factor for each target temperature. Consequently, in practice, the accumulated term is an array with only four indices (two electronic bands, one $\mathbf{k}$-point, and one temperature), explicitly avoiding a five-dimensional array that would require storing the $\mathbf{q}$ index in memory. Finally, we do the sum over $n$, $n'$ and $\mbf{k}$ to produce the density as a function of temperature. The Fourier transform gives the structure factor.\\

\subsection{Reciprocal space}

An alternative and less numerically demanding approach actually avoids calculating the charge density.

In the reciprocal case, we are interested directly in the expression after the Fourier transform. Considering the change of variables $\mathbf{\tilde{r}} = R^{-1} \mathbf{r}$ (its Jacobian is 1), we have

\begin{widetext}
\be
\begin{split}
\int \psi_{n\mbf{k}}(\tilde{\mbf{r}}) \psi^\ast_{n'\mbf{k}}(\tilde{\mbf{r}}) e^{i \mbf{K} \cdot \mbf{r}} d \mbf{r} & = \sum_{\mbf{G},\mbf{G}'} c_{n,\mbf{k}+\mbf{G}} c^\ast_{n',\mbf{k}+\mbf{G}'} \int e^{i (\mbf{G} - \mbf{G}') \cdot \tilde{\mbf{r}}} e^{i\mbf{K} \cdot \mbf{r}} d \mbf{r} \\
& = \sum_{\mbf{G},\mbf{G}'} c_{n,\mbf{k}+\mbf{G}} c^\ast_{n',\mbf{k}+\mbf{G}'} \int_{\tilde{\Omega}} e^{i (\mbf{G} - \mbf{G}') \cdot \tilde{\mbf{r}}} e^{i R'^{-1} \mbf{K} \cdot \tilde{\mbf{r}}} d \tilde{\mbf{r}}\\
& = V \sum_{\mbf{G}} c_{n,\mbf{k}+\mbf{G}} c^\ast_{n',\mbf{k}+\mbf{G}+R'^{-1} \mbf{K}} 
\end{split}
\ee

\ni Also

\be
\begin{split}
\int_\Omega e^{i \mbf{K} \cdot \mbf{r}} d\mbf{r} & = \int_{\tilde{\Omega}} e^{i \mbf{K} \cdot (\tilde{\mbf{r}} - \mbf{r}_0)} d\tilde{\mbf{r}} =  e^{- i \mbf{K} \cdot \mbf{r}_0} \int_{\tilde{\Omega}} e^{i \mbf{K} \cdot \tilde{\mbf{r}}} d\tilde{\mbf{r}} = V e^{- i \mbf{K} \cdot \mbf{r}_0} \d_{\mbf{K},0} = V \d_{\mbf{K},0}
\end{split}
\ee

\ni For the term involving the derivative, 

\be
\begin{split}
\int_\Omega \p_\b(\psi_{n\mbf{k}} \psi^\ast_{n'\mbf{k}})|_{\mbf{r}=\tilde{\mbf{r}}} e^{i \mbf{K} \cdot \mbf{r}} d \mbf{r} & = \int_{\tilde{\Omega}} \tilde{\p}_\b(\psi_{n\mbf{k}}(\tilde{\mbf{r}}) \psi^\ast_{n'\mbf{k}}(\tilde{\mbf{r}})) e^{i (R'^{-1}\mbf{K}) \cdot \tilde{\mbf{r}}} d \tilde{\mbf{r}}  \\
& = -i(R'^{-1} K)_\b V \sum_{\mbf{G}} c_{n,\mbf{k}+\mbf{G}} c^\ast_{n',\mbf{k}+\mbf{G}+R'^{-1} \mbf{K}} 
\end{split}
\ee
\end{widetext}

\ni So for both terms we have to calculate the ``shifted" coefficients, which depend on the rotation.

\section{DW factor}
\label{sec:DW}

Partitioning the charge density, we can write

\be
S_\mbf{K} = \sum_{li} f_{i\mbf{K}} e^{i \mbf{K} \cdot (\mbf{\tau}_i+\mbf{u}_{li})}
\ee

\ni where $f_{i\mbf{K}}$ is the form factor of each atom (Fourier transform of the charge density around each atom), and we used that $e^{i \mbf{K} \cdot \mbf{R}_l}=1$ for a reciprocal lattice vector $\mbf{K}$. To do the thermal average, we use that

\be
\langle e^{i \mathbf{K} \cdot \mathbf{u}_{li}}\rangle = e^{-\tfrac{1}{2} \langle (\mathbf{K} \cdot \mathbf{u}_{li})^2\rangle}
\ee

\ni Writing as usual

\be
u_{li\a} = \f{1}{\sqrt{N}} \sum_{\mbf{q}s} \f{e^{i \mbf{q} R_l}\e_{s,i\a}(\mbf{q})}{\sqrt{2 \o_{\mbf{q}s}M_i}} (a_{\mbf{q}s} + a^\dagger_{-\mbf{q}s}),
\ee

\ni we get

\be
\langle (\mbf{K} \cdot \mbf{u}_{li})^2 \rangle = \f{1}{N}\sum_{\mbf{q}s}\f{|\mbf{K} \cdot \e_{s,i}(\mbf{q})|^2}{2 M_i \o_{\mbf{q}s}} (2 n_{\mbf{q}s}+1) := W_{\mbf{K}i},
\ee

\ni where the thermal average forces the second phonon wavevector to be minus
the first, so that the phases $e^{i(\mbf{q}+\mbf{q}')\cdot \mbf{R}_l}$ cancel.
Since $W_{\mbf{K}i}$ does not depend on $l$, the sum over cells gives a factor
$N$, and per primitive cell

\be
\langle S_\mbf{K} \rangle = \sum_i f_{i\mbf{K}} e^{i \mbf{K} \cdot \mbf{\tau}_i} e^{-\f{1}{2} W_{\mbf{K}i}}.
\ee

\section{Factorization of the intensity}
\label{sec:factorization}

Using the expression for $S_\mbf{K}$, given by Eq.~\eqref{eq:SK}, the intensity in X-ray diffraction is in principle given by \cite{Coppens1997}

\be
I = \langle |S_\mbf{K}|^2 \rangle =  \int d\mbf{r} d\mbf{r}' \langle \rho(\mbf{r})\rho(\mbf{r}')\rangle e^{i \mbf{K} \cdot ( \mbf{r}-\mbf{r}')}
\label{eq:intensity_exact}
\ee

\ni To second order in atomic displacements,

\be
\begin{split}
\left[\rho(\mbf{r})\rho(\mbf{r}')\right]^{(2)} & = \rho_0(\mbf{r}) \d \rho(\mbf{r}')^{(2)} + \d \rho(\mbf{r})^{(2)} \rho_0(\mbf{r}') \\
& + \d \rho(\mbf{r})^{(1)} \d \rho(\mbf{r}')^{(1)}
\end{split}
\ee 

\ni We will now see that the thermal average of the last term is negligible, and thus Eq.~\eqref{eq:intensity_exact} reduces to Eq.~\eqref{eq:intensity} of the main text.

The expression of $\d \rho^{(1)}$ just follows from Eq.~\eqref{eq:drho_du} by multiplying by $u_{li\a}$ and summing over these indices. As usual we can write $\mbf{k}' = \mbf{k}+\mbf{q}$ because of momentum conservation. 

The product of the structure factors is

\be
\left(\int d \mbf{r} \delta \rho^{(1)} (\mbf{r}) e^{i \mbf{K} \cdot \mbf{r}}\right)  \left( \int d \mbf{r}' \delta \rho^{(1)} (\mbf{r}') e^{i \mbf{K} \cdot \mbf{r}'}\right)^\ast
\ee

Omitting the c.c., the left bracket is

\be
\sum_{nn',\mbf{k}\mbf{q}} \f{V^{(1)}_{n'\mbf{k}+\mbf{q},n\mbf{k}}}{\vare_{n\mbf{k}} - \vare_{n'\mbf{k}+\mbf{q}}} \int d\mbf{r} e^{i \mbf{K} \cdot \mbf{r}} \psi_{n'\mbf{k}+\mbf{q}} \psi_{n\mbf{k}}^\ast
\ee

\ni Using expression Eq.~\eqref{eq:psi_G} for the wavefunctions, we obtain the factor

\be
\begin{split}
\int d \mbf{r} e^{i \mbf{K} \cdot \mbf{r}} e^{i (\mbf{k} + \mbf{q} + \mbf{G}) \cdot \mbf{r}} & e^{-i(\mbf{k}+\mbf{G}') \cdot \mbf{r}} \\
 = \int d \mbf{r} & e^{i (\mbf{K} + \mbf{G} - \mbf{G}' + \mbf{q})\cdot \mbf{r}}
\end{split}
\ee

\ni and since $\mbf{K}$ and the $\mbf{G}$s are reciprocal lattice vectors, $\mbf{q}$ has to be a lattice vector as well. Since $\mbf{q}$ is in the BZ, then $\mbf{q}=0$, and this term is negligible in the thermodynamic limit for coherent scattering, where all $\mbf{q}$ contribute.

\let\underline\textit
\let\uline\textit
\let\ULon\relax
\bibliography{bibliography}

\end{document}